\documentclass[12pt,preprint]{aastex}

\begin{document}

\title{Deep Radio Continuum Imaging of the Dwarf Irregular Galaxy IC\,10: Tracing Star Formation and Magnetic Fields}

\author{V. Heesen\altaffilmark{1}, U. Rau\altaffilmark{2}, M.~P. Rupen\altaffilmark{2}, E. Brinks\altaffilmark{1}, and D.~A. Hunter\altaffilmark{3}}

\altaffiltext{1}{Centre for Astrophysics Research, University of Hertfordshire, Hatfield AL10 9AB, United Kingdom}
\altaffiltext{2}{NRAO, P.V.D.\ Science Operations Center, National Radio Astronomy Observatory, 1003 Lopezville Road, Socorro, NM 87801, USA}
\altaffiltext{3}{Lowell Observatory, 1400 West Mars Hill Road, Flagstaff, AZ 86001, USA}

\begin{abstract}
  We exploit the vastly increased sensitivity of the Expanded Very Large Array
  (EVLA) to study the radio continuum and polarization properties of the
  post--starburst, dwarf irregular galaxy IC\,10 at 6\,cm, at a linear
  resolution of $\sim50\,\rm pc$. We find close agreement between radio
  continuum and H$\alpha$ emission, from the brightest \ion{H}{2} regions to
  the weaker emission in the disk. A quantitative analysis shows a strictly
  linear correlation, where the thermal component contributes $50\%$ to the
  total radio emission, the remainder being due to a non--thermal component
  with a surprisingly steep radio spectral index of between $-0.7$ and $-1.0$
  suggesting substantial radiation losses of the cosmic--ray electrons. We
  confirm and clearly resolve polarized emission at the $10-20\%$ level
  associated with a non--thermal superbubble, where the ordered magnetic field
 is possibly enhanced due to the compression of the expanding bubble. A fraction
  of the cosmic--ray electrons has likely escaped because the measured radio
  emission is a factor of 3 lower than what is suggested by the H$\alpha$
  inferred SFR.

\end{abstract}


\section{Introduction}
Dwarf galaxies are the most abundant type of galaxy in the local Universe
\citep{mar97} and are the closest analogues to the building blocks for galaxy
formation at high redshift in $\Lambda$CDM models \cite[see, for
example,][]{lag09}. Furthermore, in the ``downsizing'' paradigm
\citep{cow95,hom06} dwarf galaxies increasingly dominate the star--forming
universe over time \citep{saw06}, with star formation (SF) perhaps being
particularly enhanced within dwarfs over the past 2 Gyr \citep{orb08}. 
Stellar winds and supernova explosions of the most massive
stars can easily lead to outflows due to their weak gravitational
potential. This implies that not only does the metallicity of the interstellar
medium increase with cosmic time, but additionally the intergalactic medium
becomes enriched at an early stage \citep{opp06}.  Moreover, dwarf galaxies
are supposed to have played a fundamental role in amplifying magnetic
fields in the early Universe through outflowing plasma escaping from
star--forming regions \citep{chy11}.

Cosmic rays play an important role in galactic outflows, as they consist of
relativistic particles that do not quickly lose their energy, like the hot
gas. They spiral along magnetic field lines emitting highly linearly polarized, 
non--thermal (synchrotron) emission providing a
useful tool to study magnetic field structure. This magnetic field
influences how cosmic rays propagate. For example, nearby edge--on
galaxies with radio haloes show a characteristic ``pinched at the waist''
structure in polarised emission (and sometimes even in total intensity)
which is a consequence of an X--shaped magnetic field \citep[e.g.,][]{tue00a,hee09b,soi11a}.
 
Resolved radio continuum studies of dwarf galaxies have been lacking thus far,
because the objects are intrinsically faint. \citet{chy03a} took advantage of
the excellent sensitivity of the 100--m Effelsberg telescope to study IC\,10
in the radio continuum, both in total intensity and polarimetry. They found
extended, diffuse emission coinciding with H$\alpha$ emission, and polarized
emission close to the non--thermal bubble described by
\citet{yan93a}. However, they were limited by their coarse resolution
($1\farcm 1$) that prevents any detailed comparison with ancillary data. A
higher resolution preliminary VLA map at 6\,cm was published by \citet{chy05}.
\citet{chy00a}, using the VLA, found in NGC\,4449 despite its weak rotation a
surprisingly regular magnetic field showing hints of spiral structure and
``fans'' possibly connected to outflows. \citet{kep10a}, again using the VLA,
studied the starburst dwarf NGC\,1569 which has a strong radial field, likely
shaped by the outflowing gas.

With the new capabilities of the expanded very large array (EVLA) and much
improved sensitivity we can study the weak
extended emission in dwarf galaxies. In this letter we present first
results from observations of IC\,10 with the EVLA at C--band, using the full
2\,GHz of bandwidth that was available at the time of observation. Additional
observations at L--, X--, and Ka--band, along with a full interpretation of
the data will be published in a forthcoming paper. 

IC\,10 is a perfect
candidate for an in depth study of a dwarf galaxy at radio continuum
wavelengths because it is relatively bright and nearby, being a member of the
Local Group \citep{san08a}. It has a high star formation rate and a high
density of Wolf--Rayet stars and is regarded to be in a
starburst phase \citep{mas02a,crow03a}. The aftermath of massive star
formation leaves its imprint on the ISM in the form of expanding bubbles and
shells that can be traced in \ion{H}{1} and H$\alpha$. \citet{wil98a} found
holes in \ion{H}{1} and H$\alpha$ emission with large filaments in H$\alpha$
extending up to a few hundred parsec away from the main body \citep{hun06a},
suggesting outflows. The \ion{H}{1} emission is asymmetric with respect to the
major axis. A feature visible in velocity maps suggests ongoing accretion of
neutral hydrogen in the outer parts of the galaxy. There is a detection of
diffuse hot X--ray emitting gas in this galaxy that is aligned with the main
stellar bar \citep{wan05a}. \citet{yan93a} found a region with high radio
continuum brightness and identified it as a non--thermal bubble prior to the
break out. \citet{loz07a} speculated that it could be generated by a hypernova
rather than by a collection of supernovae.

As IC\,10 is close to the Galactic midplane ($b=-3\fdg 3$), estimates of its distance are uncertain due to the unknown level of absorption. Here, we use a conservative distance of $1\,\rm Mpc$ in line with the most recent results that indicate a likely distance at best slightly less than 1\,Mpc \citep{kim09a,san08a}.

\section{Observations}

We observed IC\,10 with the NRAO\footnote{The National Radio Astronomy
  Observatory is a facility of the National Science Foundation operated under
  cooperative agreement by Associated Universities, Inc.} EVLA under the
resident shared risk program (RSRO; project ID: AH1006). Our observations
were taken in D-configuration on August 16, 2010. We used 2 IFs each
consisting of 8 sub--bands with a bandwidth of 128\,MHz each. The sub--bands
contain 64 channels at 2\,MHz resolution resulting in a total bandwidth of
2\,GHz. We placed our IFs between $4.5-5.4\,\rm GHz$ and $6.9-7.8\,\rm GHz$,
respectively. We observed for 4\,h to ensure good $(u,v)$--coverage and
sensitivity to extended emission. Every 15\,min we observed a phase calibrator
(J0102+5824); we used 3C48 as our flux density (primary) calibrator, adopting
the \citet{baa77} flux scale. In order to measure the polarization leakage
(D--terms) we observed the unpolarized source 3C84. The polarization angle was
calibrated using 3C48.

We calibrated the data with the Common Astronomy Software Applications package
(CASA) as outlined in the continuum tutorial of 3C391 on the CASA
homepage\footnote{http://casaguides.nrao.edu/index.php?title=EVLA\_Continuum\_Tutorial\_3C391}. In all subbands, leakages of $5\%-15\%$ before calibration were reduced to
  0.2\% (variation across channels) after calibration. R-L polarization angles
  varied by $5\degr-10\degr$ with a constant slope across channels before
  calibration, and went down to $1\degr-2\degr$ variations across channels
  after calibration. We constructed Stokes $I$, $Q$, and $U$ images using
the Multi--Scale Clean algorithm \citep{cor08a,ric08a} in CASA that models
extended emission accurately, but assumes a flat sky--spectrum.  In Stokes $I$
the integrated flux density was $131\,\rm mJy$, with a peak brightness of
$11.3\,\rm mJy$. The integrated polarized intensity is $3.0\,\rm mJy$ with
peak brightness of $-58\,\mu$Jy and $47\,\mu$Jy in Stokes $Q$ and $U$,
respectively.  The target rms noise calculated for this observation with
$\sim$3 hours on--source was $4\,\mu\rm Jy$. The rms noise in the Stokes $Q$
and $U$ images was $5\,\mu\rm Jy$, but the Stokes $I$ image was affected by
deconvolution errors at the $40\,\mu\rm Jy$ level. These errors were at the
expected level for continuum imaging when ignoring spectral structure (average
dynamic range $\sim$300 for a spectral index of $-0.7$ across a 50\%
bandwidth, $4.5-7.8\,\rm GHz$).

\begin{figure}
\resizebox{\hsize}{!}{\includegraphics{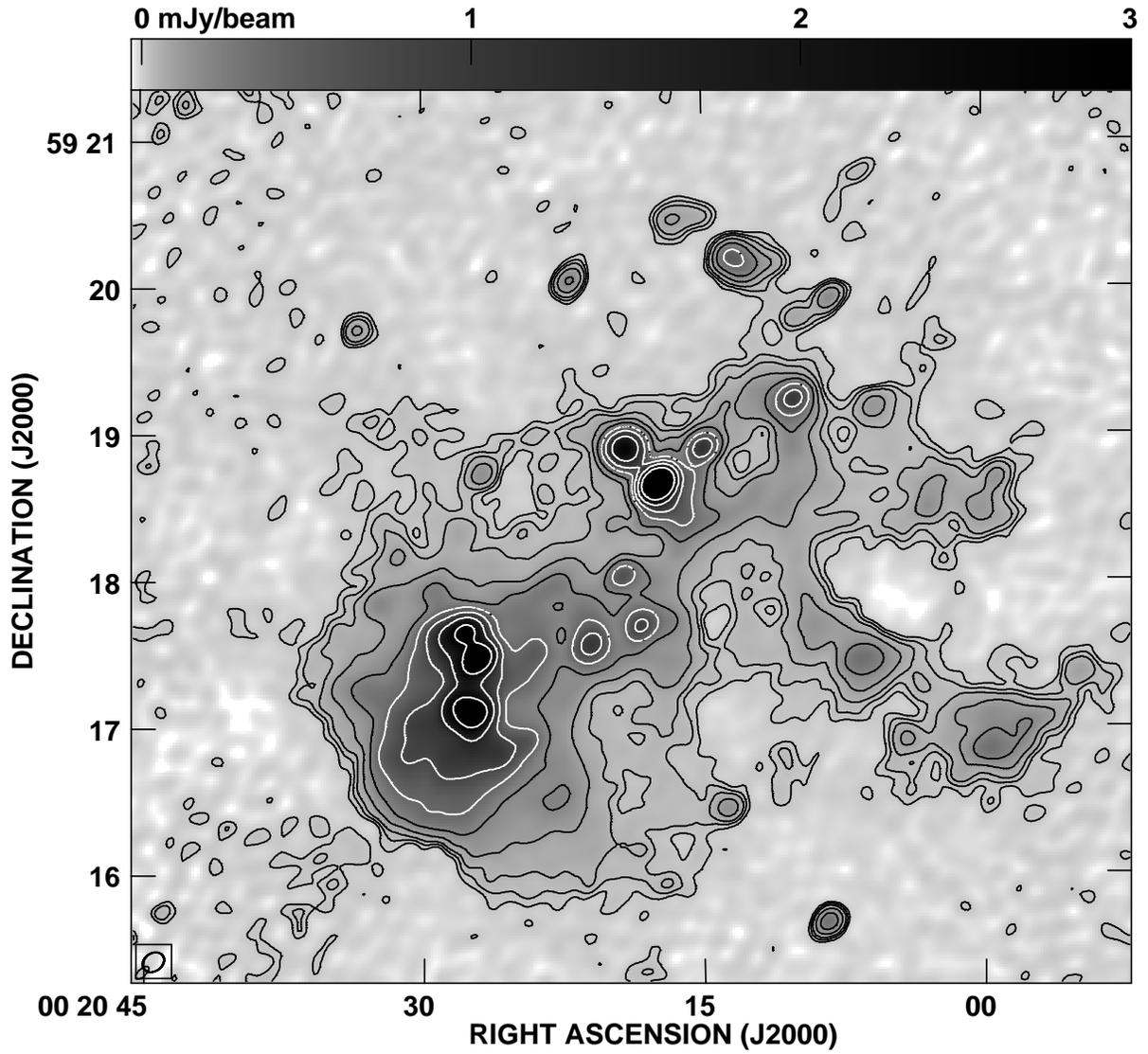}}
\caption{Total power radio continuum at 6\,cm at $9\farcs 4 \times 7\farcs 3$
  resolution as a quasi--logarithmic grey--scale image. Contours are at 3, 6,
  10, 20, 40, 80, 150, 300, and 600 $\times$ $5\,\mu\rm Jy\,beam^{-1}$. The
  rms noise level is at $5\,\mu\rm Jy\,beam^{-1}$. The total power map is
  corrected for primary beam attenuation.}
\label{fig:i}
\end{figure}

Stokes $I$ imaging was repeated with the Multi--Scale
Multi--Frequency Synthesis (MS--MFS) algorithm \citep{rau10a,rau11a} which
simultaneously solves for spatial and spectral structure during wide--band
image reconstruction, and reached an rms noise of $15\,\mu\rm Jy$.  This
wide-band model was then used to perform one round of amplitude and phase
self--calibration, and the resulting image (shown in greyscale in Fig.\,\ref{fig:i})
achieved the rms noise of $5\,\mu\rm Jy$ only 20\% higher than the theoretically expected level.
We note that the rms
is identical for Stokes $I$, $Q$, and $U$, so that we are not dominated
by the signal--to--noise ratio in Stokes $I$; the self--calibration is
sufficiently accurate.

A radio spectral index image was
produced by MS--MFS as well, and a post--deconvolution wide--band primary beam correction
was applied to remove the effect of the frequency--dependent primary beam.  
One of the goals of this RSRO project was to test the accuracy of this
approach. Therefore, this spectral index map was compared with a
low--resolution spectral index map constructed within AIPS from
primary--beam corrected Stokes $I$ images at each sub--band and smoothed to the
resolution of the lowest sub--band. The results agreed, confirming that for this
source, wide--band imaging via MS--MFS followed by a post--deconvolution
primary beam correction was able to recover the same spectral information as
traditional methods, but at a much improved angular resolution.

The Stokes $Q$ and $U$ images were taken into AIPS\footnote{AIPS, the
  Astronomical Image Processing Software, is free software from NRAO.}, and
maps of the polarized intensity were constructed using $\tt comb$ with option
$\tt polc$ that applies a correction to the noise bias. We also created the
maps of the polarization angle in comb using the option $\tt pola$. We rotate
the polarization \mbox{(E--)vectors} by $90\degr$ to show the orientation of the
magnetic field, where we neglect any contribution of Faraday rotation. We
tested whether the structure of Stokes $Q$ and $U$ changed with frequency by
making images for the 16 individual sub--bands and found the change to be small,
indicating negligible Faraday rotation within our frequency range. 
We calculated the degree of polarization and found values between
$5-20\%$. These values are at least a factor of 5 larger than what we would
expect from instrumental polarization which is around $1\%$ for the VLA beam
\citep{con98a}. This was further checked by us as we carried out separate test
observations with 3C84 at the half--power point in C--band and found values
there of at most 3\%. However, because of beam smearing and the fact that our
source lies well within the half--power point, we expect instrumental
polarization to be considerably smaller. We therefore conclude that our
measured polarization is dominated by the intrinsic polarization of our
source.

\section{Results}

Figure\,\ref{fig:i} shows our total power continuum radio map at 6\,cm with a resolution of $9\farcs 4 \times 7\farcs 3$. Contours start at 3 $\times$ the rms noise level and increase by a factor of about 2 with every further contour. We choose a logarithmic grey scale to show the noise floor that is evenly distributed over the full extent of the map. To our knowledge this is the deepest radio continuum map of a dwarf galaxy yet made, going deeper than the preliminary VLA map, also at 6\,cm, published by \citet{chy05} which reaches a noise of 9\,$\mu$Jy (Chy\.zy, priv.\ comm.). The deep VLA map of NGC\,1569 at 6\,cm by \citet{kep10a} has an rms noise of $8.5\,\mu\rm Jy\,beam^{-1}$ but at a smaller beam size of $\sim 4\arcsec$ and is much less sensitive to extended emission.  Although there are no artefacts visible in the figure that could be ascribed to missing short spacings, it can not be ruled out that we miss  low--level, extended emission. We find a total flux of 131\,mJy whereas we expect, based on interpolating single dish observations at 10.45\,GHz and 2.64\,GHz \citep{chy03a,chy11} 190\,mJy. We could thus be missing up to 30\% in total power.

\begin{figure}
\resizebox{\hsize}{!}{\includegraphics{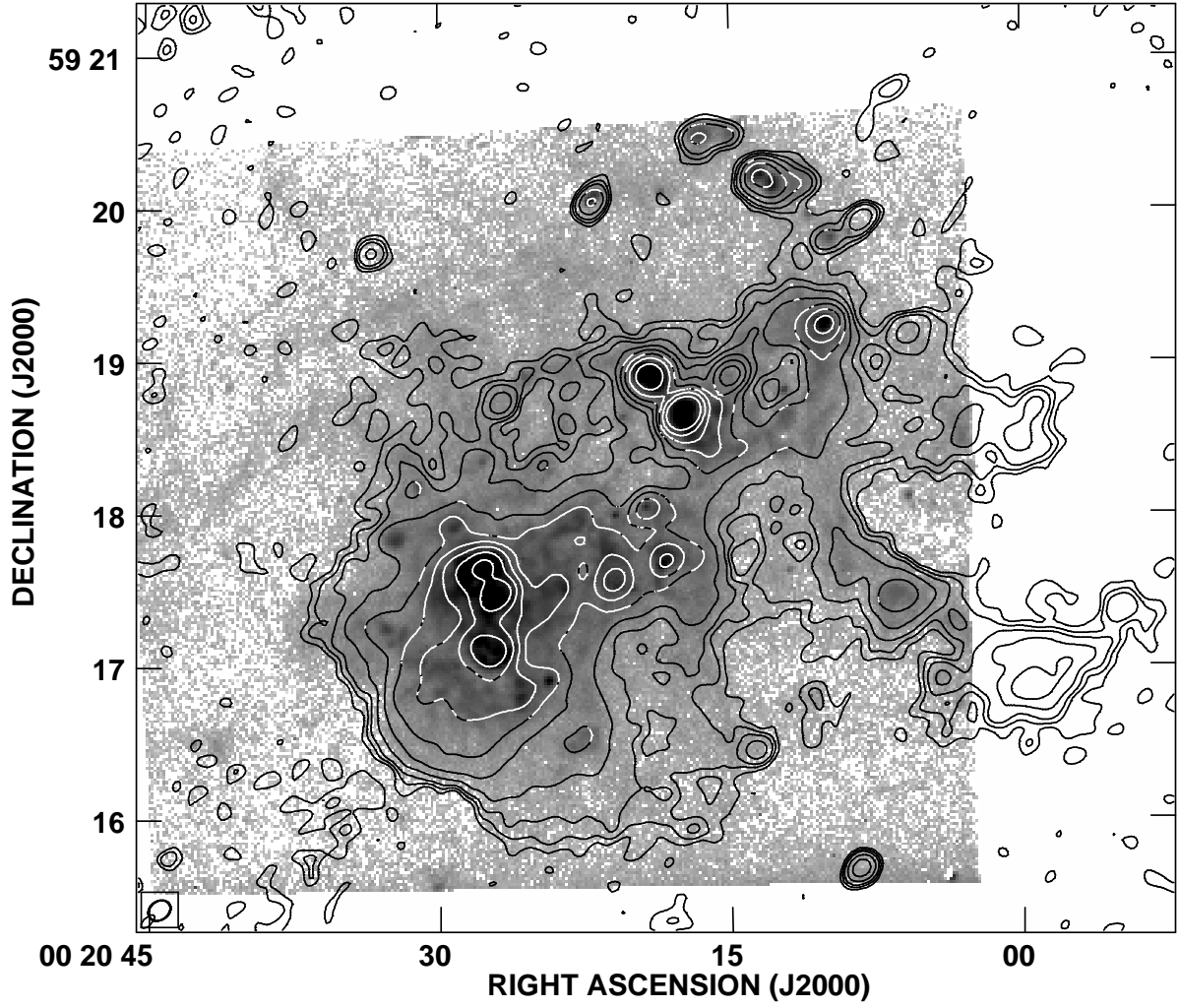}}
\caption{Same as Fig.\,\ref{fig:i} but as an overlay on H$\alpha$ shown as a quasi--logarithmic grey--scale image.}
\label{fig:ha}
\end{figure}

We compare in Fig.\,\ref{fig:ha} the radio continuum emission with a deep
H$\alpha$ map of \citet{hun04a}. The H$\alpha$ is shown as a logarithmic
grey--scale map to enhance weak emission features away from the main stellar
body. The radio and H$\alpha$ emission follow each other closely.  The radio
continuum peaks at the location of the H$\alpha$ emission maxima that trace
\ion{H}{2} complexes. Moreover, thanks to the exquisite dynamic range, we
  can compare the radio emission with extended H$\alpha$ emission, including various shells and filaments. The large filament in the West with a length
of $600\,\rm pc$ is the most prominent feature but we are sensitive enough to
also detect the much weaker filaments in the Eastern half. The most prominent
hole in the radio continuum is found at RA = $00^{\rm h}20^{\rm m} 23^{\rm
  s}$, DEC = $59\degr 18\farcm 4$ and is fully contained, both in H$\alpha$
and radio. It was identified by \citet{wil98a} to be the most prominent hole
in \ion{H}{1} emission with hints of expansion.

\begin{figure}
\resizebox{\hsize}{!}{\includegraphics{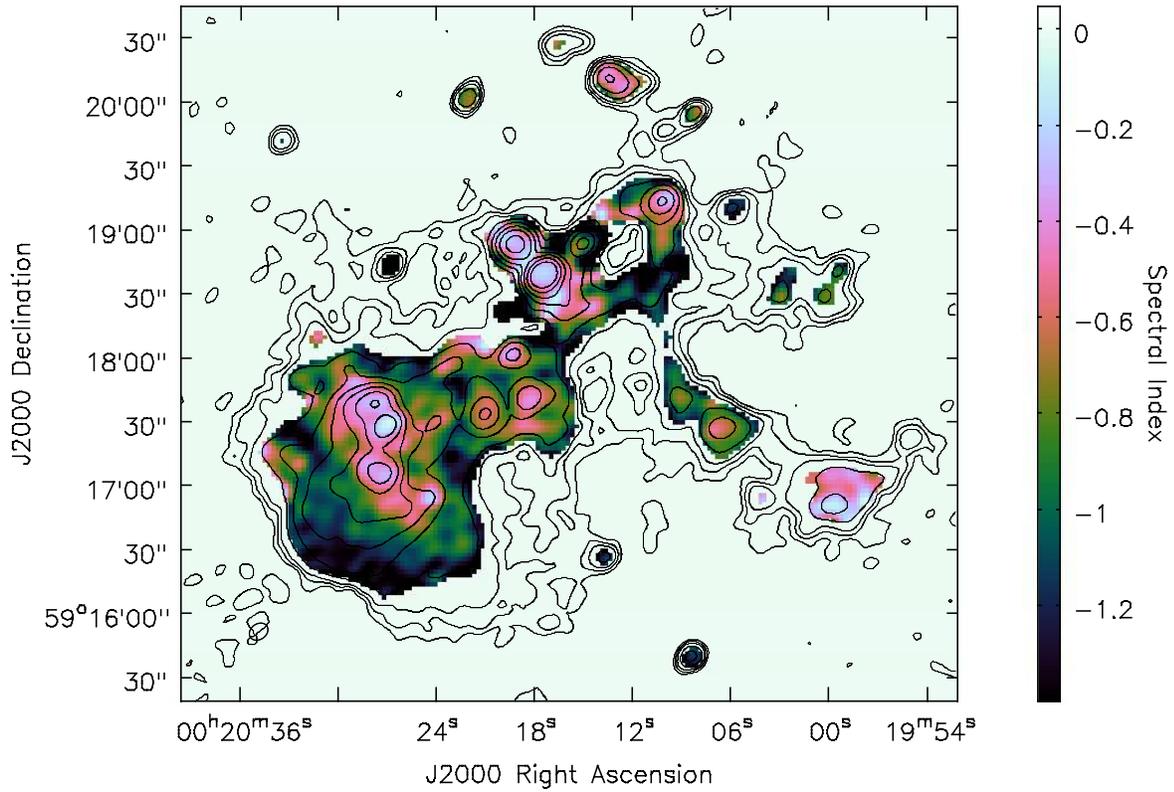}}
\caption{Radio spectral index map shown as a color--coded image at $9\farcs 4
  \times 7\farcs 3$ resolution. The contours show the radio continuum emission
  and are similar to those in Fig.\,\ref{fig:i}. The frequency dependence of
  the primary beam was removed in a post--imaging step.}
\label{fig:spix}
\end{figure}

The close correlation between the radio continuum and H$\alpha$ emission suggests a large fraction of thermal radio emission from free--free emission. This is borne out by the radio spectral index map presented in Fig.\,\ref{fig:spix}. The thermal emission has a radio spectral index of $\alpha=-0.1$ ($S\propto \nu^\alpha$), whereas the non--thermal synchrotron emission has a spectral index of $-0.7$ (provided the cosmic ray electrons are young, otherwise the spectrum is yet steeper). We find that almost all compact resolved sources within the galaxy have a flat spectral index between $-0.1$ and $-0.2$ indicative of a dominant thermal component. These are the compact \ion{H}{2} regions with intense star formation visible in H$\alpha$ emission. We note that our radio spectral index map obtained via MS--MFS has a better resolution than that obtained with the conventional method, where the resolution is limited by the low frequency map. This allows us to reliably identify the compact sources in the radio spectral index map that otherwise would be confused in the lower resolution version.

Away from the \ion{H}{2} regions the radio spectral index steepens to values
between $-0.7$ and $-1$. This can only be explained with a non--thermal
synchrotron component. Also, this requires synchrotron and inverse Compton
losses to be important for the cosmic--ray electrons as the radio spectral
index is significantly steeper than $-0.7$, the value expected for young
cosmic--ray electrons that have been accelerated in supernova shock
waves. This result is surprising as cosmic rays in starburst dwarf galaxies
are supposed to be advected in the outflow and their residence time would be
short compared to their cooling time. The radio spectral index steepens even
more dramatically at the south--eastern edge (RA = $00^{\rm h}20^{\rm m}
  30^{\rm s}$,  DEC = $59\degr 16\farcm 5$) where the superbubble of
\citet{yan93a} resides. There the values are between $-1$ and $-1.2$
indicating dominating synchrotron and inverse Compton losses. 

\begin{figure}
\resizebox{\hsize}{!}{\includegraphics{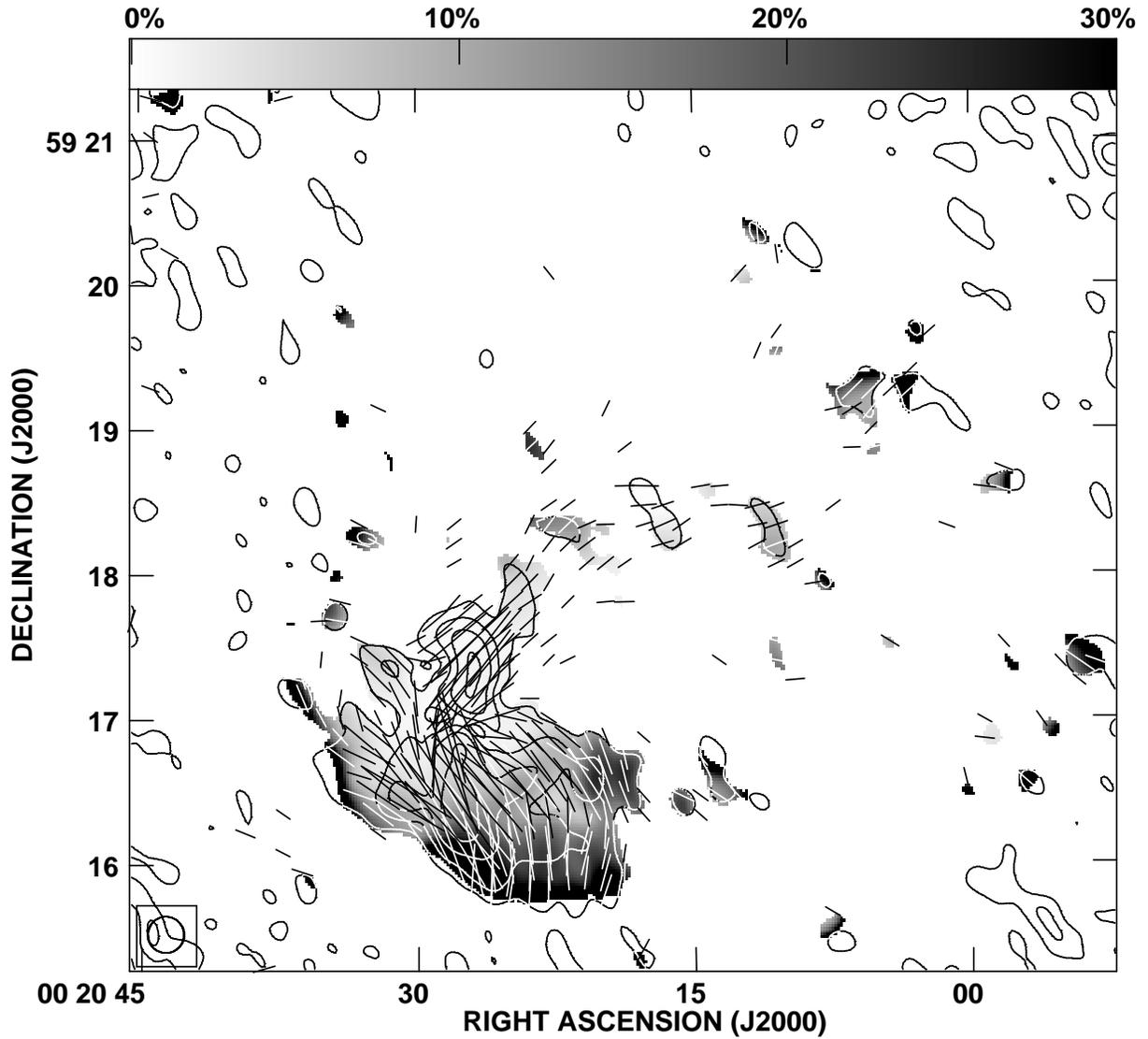}}
\caption{Polarized intensity and magnetic field orientation overlaid on
    the fractional polarization (grey scale) at $15\arcsec$
    resolution. Polarized intensity contours are at 3, 6, 10, and 20 $\times$ $7\,\mu\rm
    Jy\,beam^{-1}$. The vector length is proportional to the polarized
    intensity, where $10\arcsec$ is equivalent to $22\,\mu\rm Jy\,beam^{-1}$
    (corrected for primary beam attenuation).}
\label{fig:ha_B-vec}
\end{figure}

The map of the structure of the regular magnetic field is presented in
Fig.\,\ref{fig:ha_B-vec}. We show the orientation of the large--scale magnetic
field as measured from the linear polarization, where the length of the
vectors is proportional to the polarized intensity. The degree of polarization
in this region is $10-20\%$ indicating a significant ordered magnetic field
component possibly enhanced by compression by the expanding superbubble. The
magnetic field orientation is approximately NE-–SW, where the polarized
emission extends to the southern tip of the galaxy.

We also detect a magnetic field at the location of the two brightest \ion{H}{2}
regions in the southeastern part.  At this position, away from the
superbubble, the magnetic field orientation is more aligned with the stellar
body. We can not decide, based on the current data, if the magnetic field traverses the \ion{H}{2}
regions, because the field could also lie in front of or behind them, in a thick gas
layer as is expected for dwarf galaxies.  At this resolution we do not
detect a magnetic field that extends the full extent of the galaxy.  We note
that the location of our polarized intensity and the orientation of the
magnetic field agrees well with that measured with the 100--m Effelsberg
telescope \citep{chy03a} and the map of \citet{chy05}, lending confidence to the performance of the EVLA and the CASA data reduction package.

\section{Discussion}

\begin{figure}
\resizebox{\hsize}{!}{\includegraphics{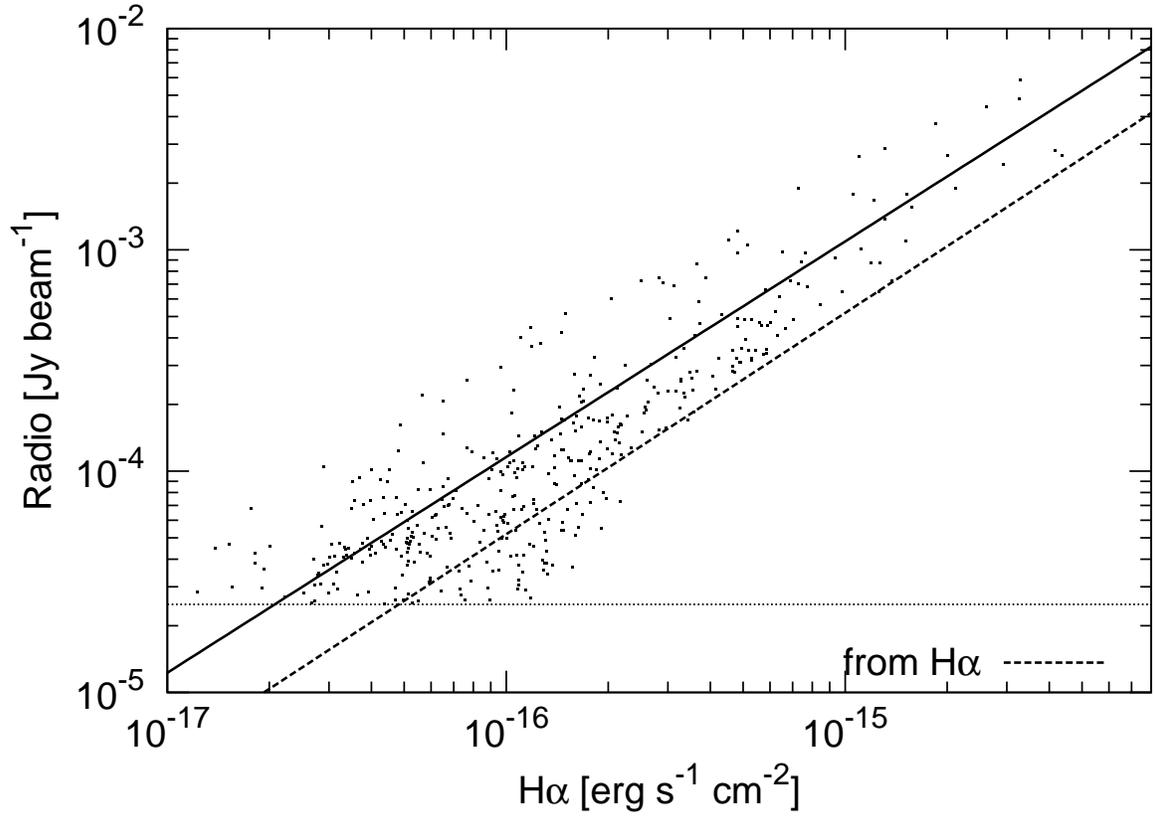}}
\caption{Comparison of the flux densities of the radio continuum emission at
  6\,cm with that of H$\alpha$ at a scale of $50\,\rm pc$. The solid line
  shows a fit to the data as described in the text. The dashed line
  shows the expected contribution based on H$\alpha$ to the radio continuum via thermal free-free
  emission (see text for details). The horizontal line indicates 5 $\times$
  the rms noise level in the radio map, where we clipped the radio data.
}
\label{fig:flux}
\end{figure}

We compared the flux densities of the radio emission and H$\alpha$ emission
averaged in $50\, \rm pc$ ($10\arcsec$) boxes. The radio and H$\alpha$ maps
were convolved with a Gaussian to a HPBW of $50\,\rm pc$ prior to
averaging. The result is presented in Fig.\,\ref{fig:flux}. The H$\alpha$ map
was corrected for foreground absorption using $\rm E(B-V) = 0.75$
\citep{bur84a}. We clipped the radio data below $5$ $\times$ the rms
  noise level, to ensure we only use reliable data points and to moderate the influence
  of missing flux density on our interferometric observations. A power law
fit in the double logarithmic plot results in a power law index of $0.97\pm
0.03$ and is therefore consistent with a strictly linear relation between the
two variables. The dashed line shows the relation if the radio flux density
were purely thermal free--free emission calculated from the H$\alpha$ flux
density \citep[e.g.][Eq.\,3]{dee97a}, assuming an electron temperature of
$10^4\,\rm K$. The thermal flux density alone is a factor 2 lower than the
measured radio flux density. This is consistent with a thermal fraction of
50\% in this galaxy as indicated by the flat radio spectral index.

The linear relation between the radio and the H$\alpha$ brightness means that
we can use both equally well as star formation tracers. The star formation rate (SFR) from the radio continuum is given by
\citet{con92a} and can be converted to a star formation
rate density. If we convert the H$\alpha$ emission into a star formation rate
by the relation given in \citet{ken98a}, we can compare the two measurements. The radio star formation rate is a factor 
of 3 lower than that implied by
H$\alpha$. This means that IC\,10 lies beneath the radio--SFR correlation
and is radio dim. Further studies are required to investigate how this corresponds to the thermal and
non--thermal fraction of the radio emission. In theory we expect that the high frequency observations are a better tracer for the thermal component, especially since they require no correction for internal or foreground extinction. One goal of this project is to establish how well we can
use radio emission as star formation tracer even in such extreme cases as starburst dwarf galaxies.

  \citet{ler05a} found that the molecular depletion time is short in IC\,10
  and that the inferred SFRs are too high for the molecular gas content. This
  may be explained by a starburst although it
  is not clear how much gas is hidden and not traced by the CO sub--mm
  rotational lines \citep{mad97a,bol00a}. As in many dwarf galaxies the knowledge
  of the molecular gas content is still sparse, the development of an
  alternative SFR tracer becomes even more important than for normal spiral
  galaxies. 

  The magnetic field structure in this galaxy lacks a strong large--scale
  component that would be apparent in radio continuum polarization. This fits
  the expectation that the standard $\alpha\Omega$-dynamo needs shear to
  operate as near--solid body rotation will provide only weak amplification
  \citep[e.g.][]{wid02a}. Such conditions favour studies of local magnetic field
  amplification mechanisms that are normally difficult to observe in spiral
  galaxies, as in these systems shear--enhanced disk magnetic fields
  dominate. The non--thermal superbubble in IC\,10 offers the possibility to
  study what role cosmic rays and magnetic fields play in the disk--halo
  interface in porous interstellar medium conditions. State of the art MHD
  simulations by \citet{avi05a} suggest that the magnetic field gets swept up
  in the walls of the shells but can not prevent a superbubble
  break--out. Eventually the full set of observations will allow us to get a
  handle on the energetics of the superbubble and put these models to the
  test.

\acknowledgements V.H.\ is funded by the Science and
Technology Facilities Council (STFC) via a rolling grant to the Centre for
Astrophysics Research. V.H.\ gratefully acknowledges the hospitality of NRAO,
as provided under the RSRO scheme. D.A.H.\ is grateful for funding provided by the
National Science Foundation through grant AST-0707563.

\end{document}